\documentclass[aps,prd,superscriptaddress,nofootinbib,floats,preprint,floats,floatfix]{revtex4-1}
\usepackage{bm}
\usepackage{indentfirst}
\usepackage{amsmath}
\usepackage{graphicx}
\usepackage{float}
\usepackage{amssymb}
\usepackage{subfigure}
\usepackage{psfrag}
\usepackage{hyperref}
\usepackage{tensor}
\usepackage{braket}
\usepackage{multirow}
\usepackage{verbatim}
\newcommand{\tabincell}[2]{\begin{tabular}{@{}#1@{}}#2\end{tabular}}
\usepackage[dvipsnames, svgnames, x11names]{xcolor}
\hypersetup{
	colorlinks=true,
	linkcolor=red,
	citecolor=blue,
} 

\allowdisplaybreaks[1]

\begin{document}
\title{Primordial black holes and scalar induced gravitational waves from the $E$ model with a Gauss-Bonnet term}

\author{Fengge Zhang}
\email{zhangfg5@mail.sysu.edu.cn}
\affiliation{School of Physics and Astronomy, Sun Yat-sen University, Zhuhai 519088, China}

\begin{abstract}
{We study an inflationary $E$ model with the Gauss-Bonnet coupling, which can enhance the curvature perturbation at small scales and thus produce a significant abundance of primordial black holes (PBHs) and detectable scalar induced gravitational waves (SIGWs).
PBHs from the $E$ model with mass $30M_{\odot}$, $10^{-5}M_{\odot}$, and $10^{-12}M_{\odot}$ can explain the LIGO-Virgo events, the ultrashort-timescale microlensing events in the OGLE data, and all dark matter, respectively.
SIGWs produced by the $E$ model can account for the recent NANOGrav signal.
We also compute the primordial non-Gaussianity and discuss its impact on PBHs and SIGWs. 
The probability distribution of density contrast is modified to be right-tailed, which we find prompts the formation of PBHs, so that the abundance of PBHs is underestimated with Gaussian approximation.
On the contrary, the fractional energy density of SIGWs is hardly affected.}
\end{abstract}

\maketitle

\section{Introduction}
The primordial black hole (PBH) \cite{Carr:1974nx,Hawking:1971ei} dark matter (DM) has revived the research interests of the community since the detection of gravitational waves (GWs) by the Laser Interferometer Gravitational-Wave Observatory (LIGO) scientific collaboration and Virgo collaboration \cite{Abbott:2016nmj,Abbott:2016blz,Abbott:2017gyy,TheLIGOScientific:2017qsa,Abbott:2017oio,Abbott:2017vtc,LIGOScientific:2018mvr,Abbott:2020khf,Abbott:2020uma,LIGOScientific:2020stg}, as the GW events may originate from PBHs \cite{Bird:2016dcv,Sasaki:2016jop,Takhistov:2020vxs,DeLuca:2020sae,Abbott:2020niy}. 

The gravitational collapse of the overdense region in the early Universe, especially in the radiation-dominated era, is one of the most accepted mechanisms to produce PBHs.
The large density perturbations generated during inflationary era induce the gravitational collapse after horizon reentry. 
In addition to PBHs, these perturbations also induce the generation of scalar induced gravitational waves (SIGWs) that contribute to the stochastic gravitational wave background (SGWB) \cite{Saito:2008jc,Orlofsky:2016vbd,Nakama:2016gzw,Wang:2016ana,Cai:2018dig,Kohri:2018awv,Espinosa:2018eve,Kuroyanagi:2018csn,Domenech:2019quo,Fumagalli:2020nvq,Domenech:2020kqm,Domenech:2021ztg,Wang:2021djr,Adshead:2021hnm,Ahmed:2021ucx}. 
In order to produce PBHs occupying a significant portion of dark matter, the power spectrum of curvature perturbations should be enhanced to $\mathcal P_{\mathcal R}\sim \mathcal{O}(10^{-2})$ on small scales \cite{Motohashi:2017kbs,Sato-Polito:2019hws,Lu:2019sti,Khlopov:2008qy}, seven-order larger than the cosmic microwave background (CMB) constraint, that is, $\mathcal P_{\mathcal R}\sim\mathcal{O}(10^{-9})$ on large scales \cite{Akrami:2018odb}.

Some inflationary models, such as the canonical single-field inflation with an inflection point \cite{Germani:2017bcs,Gong:2017qlj,Garcia-Bellido:2017mdw,Xu:2019bdp}, or a steplike feature \cite{Inomata:2021tpx} in potential can realize the enhancement of power spectrum.
Besides, a single-field inflaton with a noncanonical kinetic term that contains a peaked coupling function is also an efficient way to amplify the power spectrum \cite{Lin:2020goi,Yi:2020cut,Yi:2020kmq,Gao:2020tsa,Lin:2021vwc,Zhang:2020uek,Zhang:2021vak}. 
Recently, the two-field inflation models that predict PBH dark matter and observable SIGWs also attract much attention \cite{Pi:2017gih,Braglia:2020eai,Cheong:2019vzl,Gundhi:2020kzm,Palma:2020ejf,Spanos:2021hpk,Cai:2021wzd}. 

Usually, the perturbation is assumed to be Gaussian in computing the abundance of PBHs.
However, if non-Gaussianity parameter  satisfies $f_{\mathrm{NL}}\gtrsim \mathcal{O}(0.01)$, then even an unobservable non-Gaussianity could result in a large correction to the Gaussian-predicted PBH abundance \cite{Zhang:2021vak}.
Several canonical single-field inflation models with an inflection point in the inflaton potential that serves the formation of PBHs have been reanalyzed \cite{Atal:2018neu}.
It was found that the effect of non-Gaussianity of primordial curvature perturbations on PBH abundance is non-negligible in these models. 
Some noncanonical inflationary models with minimal or nonminimal coupling between gravity and inflaton also produce non-Gaussianity that has a significant effect on PBH abundance \cite{Zhang:2021vak,Lin:2021vwc}.
In fact, these noncanonical inflation models are equivalent to canonical inflation models with an inflection point in potential by a conformal transformation and field redefinition \cite{Zhang:2021vak}.
Pedagogically, it is worth studying the inflationary models whose enhancement mechanism is different from that of the inflection-point inflations, and further discussing the effect of non-Gaussianity. 

It is expected that the higher-order curvature correction of gravity plays an important role in the early Universe, for example, in the inflationary era. So, it is necessary to conceive an inflation model with higher-order curvature terms that give PBH dark matter and SIGWs.
Based on this motivation, we consider an inflationary $E$ model with the Gauss-Bonnet term.
The enhancement mechanism in this model is quite different from the inflection-point mechanism.
Inflation with the Gauss-Bonnet term was widely studied \cite{Kawai:1999pw,Satoh:2008ck,Bamba:2014zoa,Bamba:2014mya,Nozari:2016xgd,Nozari:2017rta,Yi:2018gse,Kleidis:2019ywv,Pozdeeva:2020shl,Oikonomou:2020sij,Odintsov:2020xji}.
It belongs to a subclass of the Horndeski theory, the most general scalar-tensor theory with the no-more-than-second-order equations of motion of both the metric and the scalar field \cite{horndeski1974second,Kobayashi:2011nu}. 
It is interesting to see that if the Gauss-Bonnet term can lead to an amplification of the power spectrum of curvature perturbation.
In Ref. \cite{Kawai:2021edk}, the authors use the natural potential to drive inflation with the Gauss-Bonnet term, and compute the PBH abundance with Gaussian approximation.
As mentioned above, the PBH abundance is much sensitive to non-Gaussianity \cite{Zhang:2021vak}, it is necessary to compute the non-Gaussianity and discuss its contribution to PBH abundance and corresponding SIGWs.

In this paper, we study the inflationary $E$ model with the Gauss-Bonnet coupling and compute the predicted PBH abundance and the fractional energy density of SIGWs.
We also discuss the role that primordial non-Gaussianity plays in the generation of PBHs and SIGWs.
This paper is organized as follows. 
In Sec. \ref{GB}, we compute the power spectrum from our model. 
In Sec. \ref{NG}, we compute the bispectrum and the corresponding non-Gaussianity parameter $f_{\mathrm{NL}}$.
In Sec. \ref{PBHGW}, we compute the PBH abundance and the fractional energy density of SIGWs with consideration of non-Gaussianity of primordial curvature perturbations.
Finally, we conclude this paper in Sec. \ref{conclusion}. 

\section{The Gauss-Bonnet inflation}\label{GB}
The action of the inflation model with the Gauss-Bonnet term is
\begin{equation}
    S=\frac{1}{2}\int d^4x \sqrt{-g}\left[R-g^{\mu\nu}\partial_\mu \phi \partial_\nu \phi -2V(\phi)-\frac{1}{8}f(\phi)\mathcal{L}_{GB}\right],
\end{equation}
where $\mathcal{L}_{GB}=R^2-4R_{\mu\nu}R^{\mu\nu}+R_{\mu\nu\rho\lambda}R^{\mu\nu\rho\lambda}$ is the Gauss-Bonnet term and $f(\phi)$ is a coupling function of inflaton $\phi$.

Working in the spatially flat Friedmann-Robertson-Walker metric and varying the action with respect to the metric $g_{\mu\nu}$ and the inflaton $\phi$, we get the following equations of motion
\begin{gather}\label{BEoMs}
6 H^{2}=\dot{\phi}^{2}+2 V+3\dot{f} H^{3}, \\
2 \dot{H}=-\dot{\phi}^{2}+\frac{1}{2}\ddot{f} H^{2}+\frac{1}{2} \dot{f} H\left(2 \dot{H}-H^{2}\right), \\
\ddot{\phi}+3 H \dot{\phi}+V_{\phi}+\frac{3}{2} f_{\phi} H^{2}\left(\dot{H}+H^{2}\right)=0,
\end{gather}
where the dot denotes the derivative with respect to $t$, and $f_\phi=df/d\phi$, $V_\phi=dV/d\phi$. 

Impose the slow roll conditions, $\dot{\phi}^2 \ll V(\phi)$, $H|\dot{f}|\ll 1$ and $|\ddot{f}| \ll H|\dot{f}|$, and the equation of motion for $\phi$ becomes
\begin{equation}\label{eqphi1}
    \ddot{\phi}+3 H \dot{\phi}+V_{\phi}+\frac{1}{6} V^2f_{\phi} \simeq 0.
\end{equation}
Conventionally, we assume that the acceleration of inflaton $\phi$ satisfies $-\ddot{\phi}/H\dot{\phi}\ll 1$ so that inflation lasts long enough. 
If the last two terms in Eq. \eqref{eqphi1} satisfy $V_{\phi}+ V^2f_{\phi}/6\Big|_{\phi=\phi_{c}}\simeq 0$ at some energy scale $\phi_c$, then Eq. \eqref{eqphi1} becomes
\begin{equation}
    \ddot{\phi}+3 H \dot{\phi}\simeq 0,
\end{equation}
which means the inflaton evolves into a transitory ultraslow-roll (USR) phase where the curvature perturbations is enhanced. 
This critical value $\phi_c$ is a nontrivial fixed point \cite{Kawai:2021bye}. 
So, this provides a way to adjust the parameters to realize an USR phase on different energy scales, which corresponds to an enhancement of power spectrum at different scales.

In this paper, we choose the coupling function to be \cite{Kawai:2021edk}
\begin{equation*}
    f(\phi)=f_0\tanh[f_1(\phi-\phi_c)],
\end{equation*}
and the $E$ model potential
\begin{equation}
    V(\phi)=V_0 \left[1-\mathrm{exp}{\left(-\sqrt{\frac{2}{3\alpha}}\phi\right)}\right]^p,
\end{equation}
where $\alpha$ and $p$ are two free parameters.
For convenience, we define the Hubble flow parameters
\begin{equation}
    \epsilon_1=-\frac{\dot{H}}{H^2},\ \ \epsilon_2=\frac{d\ln \epsilon_1}{Hdt},
\end{equation}
similarly, for $f(\phi)$, we have
\begin{equation}
    \xi_1=H\dot{f},\ \ \xi_2=\frac{d\ln \xi_1}{Hdt}.
\end{equation}
The quadratic action that determines the evolution of comoving curvature perturbation is \cite{DeFelice:2011zh}
\begin{equation}
    S_2=\int dt d^3x a^3Q \left(\dot{\mathcal{R}}^2-\frac{c^2_s}{a^2}(\partial\mathcal{R})^2\right),
\end{equation}
where
\begin{equation}
Q=\frac{1}{2}\left(\frac{1-\xi_1/2}{1-3 \xi_{1} / 4}\right)^{2}
\left(2 \epsilon_{1}-\frac{1}{2} \xi_{1}+\frac{1}{2} \xi_{1} \xi_{2}-\frac{1}{2} \xi_{1} \epsilon_{1}+\frac{3}{4} \frac{\xi_{1}^{2}}{2-\xi_{1}}\right),
\end{equation}
and 
\begin{equation}
c^2_s= 1-\frac{a^{2}}{z^{2}}\left(\frac{\xi_{1}}{2-3 \xi_{1} / 2}\right)^{2}\left(2 \epsilon_{1}+\frac{1}{4} \xi_{1}-\frac{1}{4} \xi_{1} \xi_{2}-\frac{5}{4} \xi_{1} \epsilon_{1}\right),
\end{equation}
with $z^2=2a^2Q$. The equation of motion for comoving curvature perturbation $\mathcal{R}$ reads
\begin{equation}\label{MSEQ}
    v^{''}_k+\left(c^2_s k^2-\frac{z^{''}}{z}\right)v_k=0,
\end{equation}
where $v_k=z\mathcal{R}_k$ and $\mathcal{R}_k$ is the mode function. The prime denotes the derivative with respect to conformal time $\tau$.

\begin{figure}[htp]
\centering
\subfigure{\includegraphics[width=0.45\linewidth]{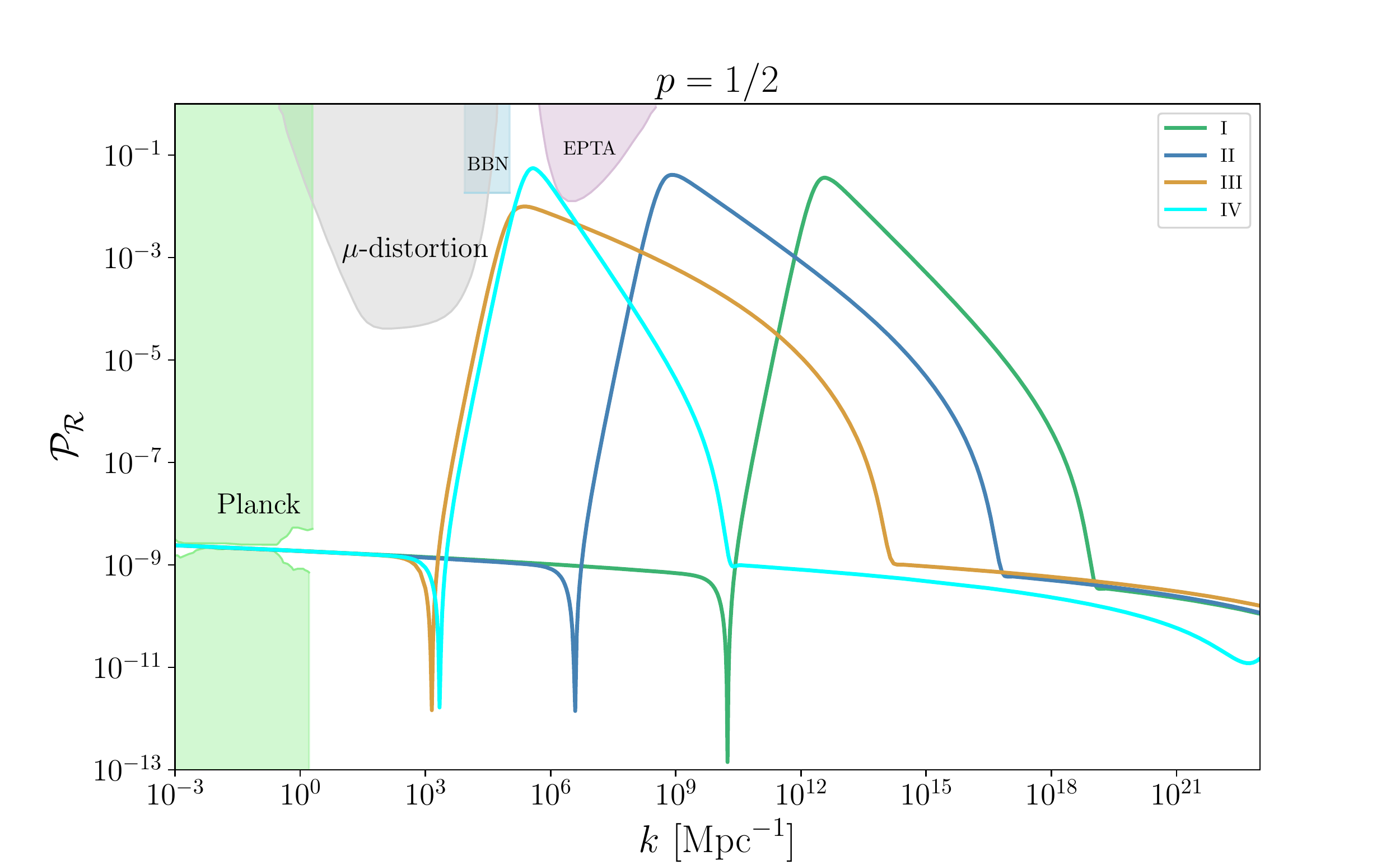}}
\subfigure{\includegraphics[width=0.45\linewidth]{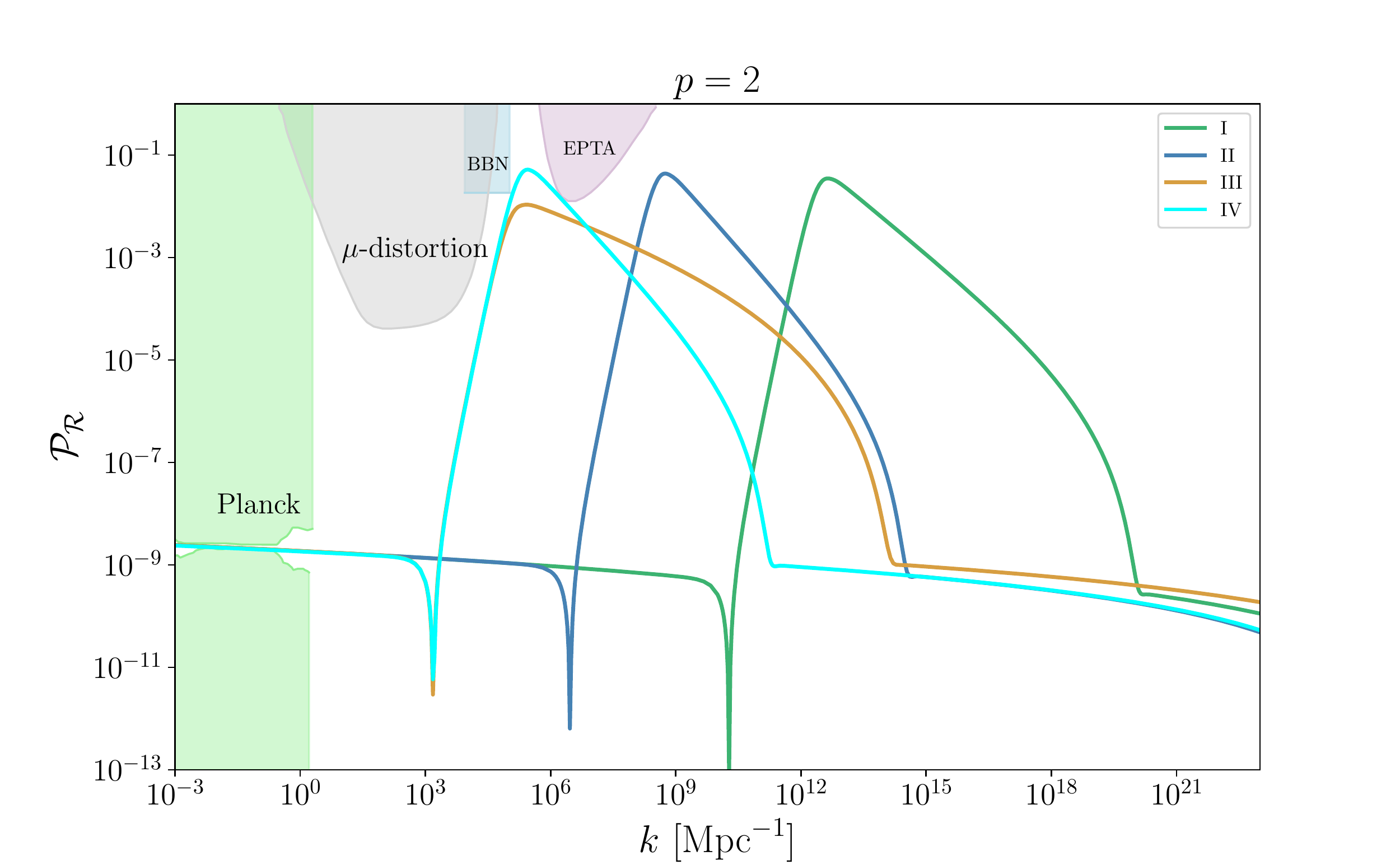}}
\caption{The power spectra produced by the $E$ model for $p=1/2$ (left panel) and $p=2$ (right panel). They are enhanced to $\mathcal{O}(0.01)$ on small scales.
The solid colored lines represent the power spectra and the shaded regions are constraints from observations \cite{Akrami:2018odb,Inomata:2018epa,Inomata:2016uip,Fixsen:1996nj}.}
\label{fig:ps}
\end{figure}

The initial condition is given by Bunch-Davies vacuum.
We numerically solve the equation \eqref{MSEQ} with the parameter sets in Table \ref{bg:tab}, and the results are shown in Tables \ref{bg:tab}, \ref{results:tab}, and Fig. \ref{fig:ps}.
From Table \ref{bg:tab} and Fig. \ref{fig:ps}, the power spectra from the $E$ model satisfy the  constraints on CMB \cite{Akrami:2018odb,BICEP:2021xfz} and the total $e$-folds $N_*\sim 60$.
At small scales, the power spectra reach $\mathcal{O}(10^{-2})$ and satisfy the constraints from CMB $\mu$-distortion, big bang nucleosynthesis (BBN) and pulsar timing array (PTA) observations \cite{Inomata:2018epa,Inomata:2016uip,Fixsen:1996nj}.

\begin{table}[htp]
\centering
\renewcommand\tabcolsep{4.5pt}
\begin{tabular}{|c|c|c|c|c|c|c|c|c|} 
\hline
&$\mathrm{Sets}$& $V_0/10^{-10}$ & $f_0/10^7$ &$-f_1$ & $\phi_c$ & $N_*$ & $n_s$ & $r$ \\
\hline \multirow{4}{*} {\tabincell{c}{$\alpha=3$ \\ $p=\frac{1}{2}$}} &$\mathrm{I}$& $2.6$ & $4.8$ & $39.302$ & $3.1$ & $63$ & $0.966$ &$0.008$\\
\cline {2-9} & $\mathrm{II}$& $3$ & $3$ & $44.47$ & $3.4$ & $62.5$ & $0.963$&$0.009$\\
\cline {2-9} & $\mathrm{III}$& $3.1$ & $1.9$ & $52.67$ & $3.8$ & $64$ & $0.963$&$0.01$\\
\cline {2-9} & $\mathrm{IV}$& $3$ & $1.65$ & $63.5323$ & $3.8$ & $55.2$ & $0.963 $&$0.009$\\
\hline \multirow{4}{*} {\tabincell{c}{$\alpha=2$ \\ $p=2$}} &$\mathrm{I}$&  $2.7$ & $6.3$ & $33.3273$ & $4.7$ & $65.7$ & $0.962 $ &$0.008$ \\
\cline {2-9} & $\mathrm{II}$& $2.7$ & $2.8$ & $50.0568$ &$5.3$ & $61.9$ & $0.962$&$0.008$ \\
\cline {2-9} & $\mathrm{III}$& $2.7 $ & $1.92 $ & $ 55.783$ &$ 5.7$ & $ 69$ & $ 0.962$&$ 0.008$ \\
\cline {2-9} & $\mathrm{IV}$& $2.6$ & $1.78$ & $62.9353$ & $5.7$ & $62.4$ & $ 0.962$ & $ 0.008$\\
\hline
\end{tabular}
\caption{The parameters of the potential and coupling function, the spectral index $n_s$, the tensor-to-scalar ratio $r$ at pivot scale $k_*=0.05\  \mathrm{Mpc}^{-1}$, and the $e$-folds $N_*$ between the horizon exit of the pivot scale modes and the end of inflation.}
\label{bg:tab}
\end{table}

\begin{table}[htp]
\centering
\renewcommand\tabcolsep{3.0pt}
\begin{tabular}{|c|c|c|c|c|c|c|c|c|} 
\hline
&$\mathrm{Sets}$&  $k_{\mathrm{peak}}/\mathrm{Mpc}^{-1}$ & $\mathcal{P}_{\mathcal{R}}/10^{-2}$ & $Y^G_{\mathrm{PBH}}$ & $M/M_{\odot}$ &$f_c/\mathrm{Hz}$&$f_{\mathrm{NL}}$ & $\Delta_3$\\
\hline \multirow{4}{*} {\tabincell{c}{$\alpha=3$ \\ $p=\frac{1}{2}$}} &$\mathrm{I}$&   $3.68\times 10^{12}$ & $3.61$ &$0.969$&$2.3\times10^{-13} $ &$6.63\times10^{-3}$& $0.34$ & $18.2$\\
\cline {2-9} & $\mathrm{II}$& $8.27\times 10^{8}$ & $4.11$&$0.0273$&$ 4.36\times10^{-6}$ &$1.48\times10^{-6}$&$0.23 $ & $11.2 $\\
\cline {2-9} & $\mathrm{III}$& $2.39\times 10^{5}$ & $0.99$&$ -$&$ -$ &$4.15\times10^{-10}$&$0.13 $ & $24.9 $\\
\cline {2-9} & $\mathrm{IV}$& $3.67\times 10^{5}$ & $5.52$&$0.0021$&$ 27.1$ &$7.02\times10^{-10} $&$0.48 $ & $ 17.1$\\
\hline \multirow{4}{*} {\tabincell{c}{$\alpha=2$ \\ $p=2$}} &$\mathrm{I}$&  $4.53\times 10^{12}$ & $3.49$ &$0.79$ &$1.58\times10^{-13} $ &$8.63\times10^{-3}$&$ 0.28$ & $ 15.6$\\
\cline {2-9} &$\mathrm{II}$&  $5.61\times10^{8}$ & $4.35$&$0.021$ & $9.98\times10^{-6} $ &$9.75\times10^{-7}$&$0.36 $ & $16.3 $\\
\cline {2-9} &$\mathrm{III}$&  $ 2.64\times10^5$ & $ 1.07$&$- $ & $ -$ &$4.53\times10^{-10} $&$0.13 $ & $ 23.8$\\
\cline {2-9} &$\mathrm{IV}$&  $3.04\times10^5$ & $5.19$&$8.8\times10^{-4}$&$ 33$ &$ 5.12\times10^{-10}$&$0.36 $ & $13.7 $\\
\hline
\end{tabular}
\caption{The table shows the peak scales, the corresponding amplitude of the power spectrum, the PBH DM abundance with Gaussian approximation, the critical frequency of SIGWs, the non-Gaussianity parameter $f_{\mathrm{NL}}(k_{\text{peak}},k_{\text{peak}},k_{\text{peak}})$ and the 3rd cumulant $\Delta_3$ at peak scales.}
\label{results:tab}
\end{table}

\section{Non-Gaussianity}\label{NG}
Given the two-point function of primordial curvature perturbations, namely, the power spectrum, we know nothing about the interaction of the primordial perturbations, as well as the non-Gaussian feature.
So it is necessary to study the higher-point correlations, in which the least order is the three-point function.
This could improve our estimation of the abundance of PBHs and the energy density of SIGWs, as the non-Gaussianity may lead to a huge modification\cite{Zhang:2021vak}.
The bispectrum $B_{\mathcal{R}}$ is related to the three-point function as \cite{Byrnes:2010ft,Ade:2015ava} 
\begin{equation}\label{Bi}
\left\langle\hat{\mathcal{R}}_{\bm{k}_{1}}\hat{\mathcal{R}}_{\bm{k}_{2}}\hat{\mathcal{R}}_{\bm{k}_{3}}\right\rangle=(2 \pi)^{3} \delta^{3}\left(\bm{k}_{1}+\bm{k}_{2}+\bm{k}_{3}\right) B_{\mathcal{R}}\left(k_{1}, k_{2}, k_{3}\right),
\end{equation}
where the three-point function $\left\langle\hat{\mathcal{R}}_{\bm{k}_{1}}\hat{\mathcal{R}}_{\bm{k}_{2}}\hat{\mathcal{R}}_{\bm{k}_{3}}\right\rangle$ can be computed using in-in formula \cite{Maldacena:2002vr,Adshead:2009cb,DeFelice:2011uc}
\begin{equation}\label{in-in}
\left\langle\hat{\mathcal{R}}_{\bm{k}_{1}}\hat{\mathcal{R}}_{\bm{k}_{2}}\hat{\mathcal{R}}_{\bm{k}_{3}}\right\rangle
\simeq 2 \Im\left[\int_{t_i}^{t_e} d t \left\langle\hat{\mathcal{R}}^I_{\bm{k}_{1}}\left(t_e\right)\hat{\mathcal{R}}^I_{\bm{k}_{2}}\left(t_e\right)\hat{\mathcal{R}}^I_{\bm{k}_{3}}\left(t_e\right) \hat{H}_{I}(t)\right\rangle\right],
\end{equation}
where $\hat{\mathcal{R}}^I$ is the operator in the interaction picture, $t_e$ and $t_i$ are the end of inflation and some early time when the perturbations are well within the horizon, respectively.
$\Im$ represents the imaginary part of the argument and the interaction Hamiltonian
\begin{equation}
    H_I=-\int d^3 x \mathcal{L}_3.
\end{equation}
The lengthy expressions of cubic Lagrangian $\mathcal{L}_3$ and the bispectrum can be found in the appendix \ref{app.NG}.
The non-Gaussianity parameter $f_\text{NL}$ is defined as \cite{Creminelli:2006rz,Byrnes:2010ft}
\begin{equation}\label{Fnl}
f_{\text{NL}}(k_1,k_2,k_3)=\frac{5}{6}\frac{B_{\mathcal{R}}(k_1,k_2,k_3)}{P_{\mathcal{R}}(k_1)
P_{\mathcal{R}}(k_2)+P_{\mathcal{R}}(k_2)P_{\mathcal{R}}(k_3)+P_{\mathcal{R}}(k_3)P_{\mathcal{R}}(k_1)},
\end{equation}
where $P_\mathcal{R}(k)=\frac{2\pi^2}{k^3}\mathcal{P}_\mathcal{R}(k)$.

\begin{figure}[htp]
\centering
\subfigure{\includegraphics[width=0.45\linewidth]{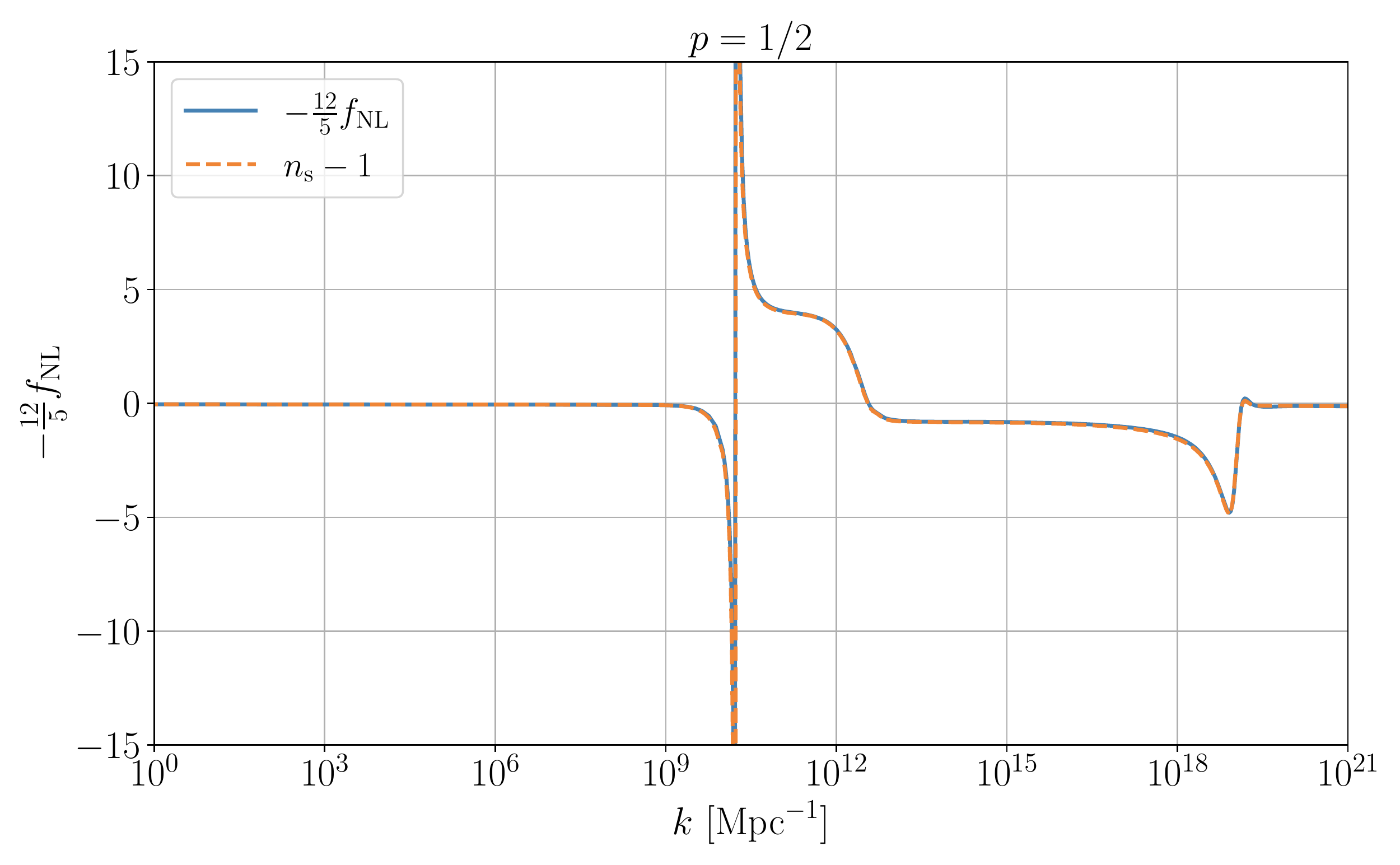}}
\subfigure{\includegraphics[width=0.45\linewidth]{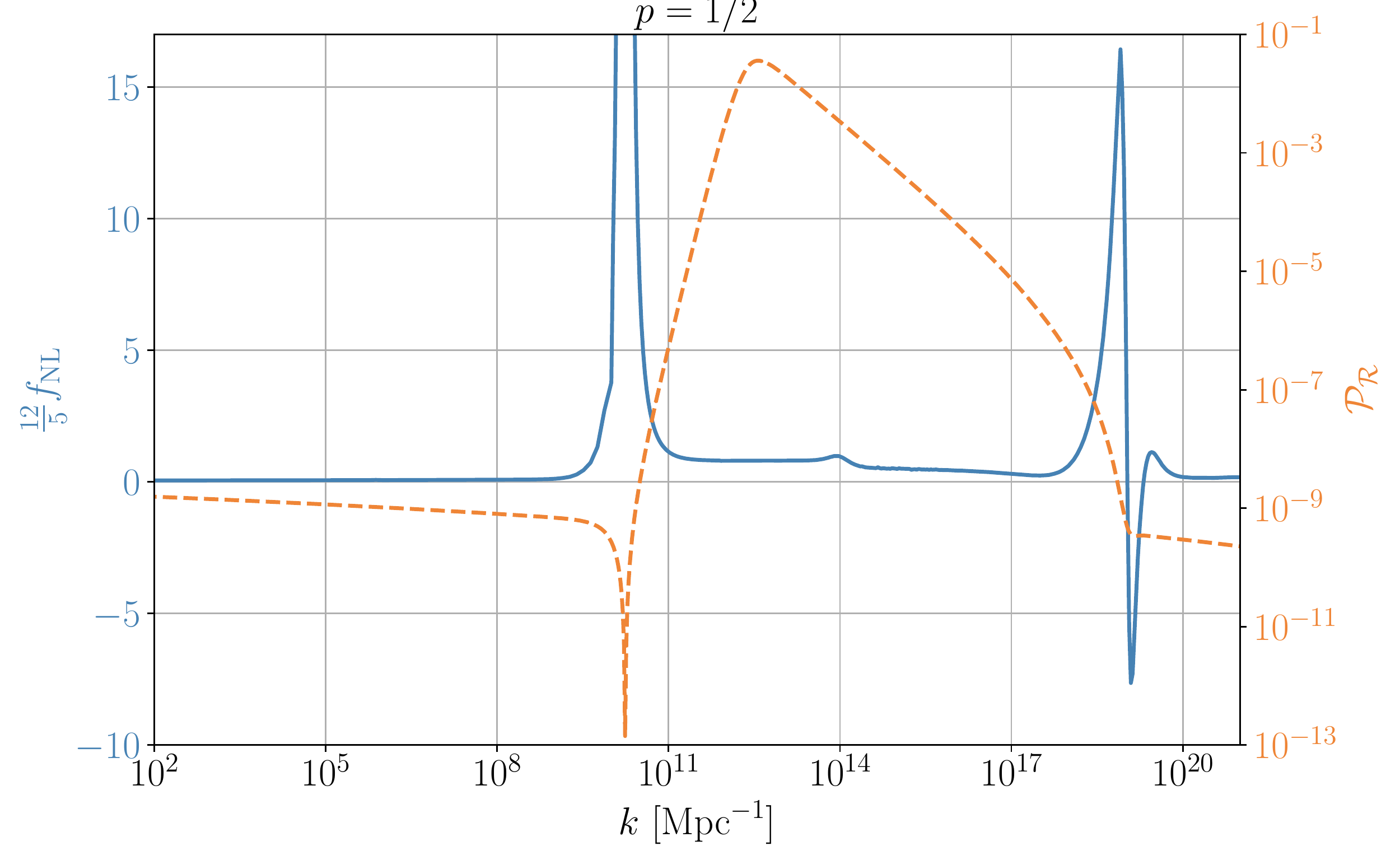}}
\caption{The non-Gaussianity parameter $f_{\mathrm{NL}}$, spectral index $n_s-1$ and power spectrum for parameter set I. In the left panel, we show $n_s-1$ in the solid line, and the non-Gaussianity parameter $-\frac{12}{5}f_{\mathrm{NL}}$ in squeezed limit for $k_1=k_2=10^6 k_3=k$ in the dashed line. In the right panel, we show the non-Gaussianity parameter $\frac{12}{5}f_{\mathrm{NL}}$ in equilateral limit for $k_1=k_2=k_3=k$ and the power spectrum in solid and dashed line, respectively.}
\label{fig:fnl}
\end{figure}

With the above formulas, we numerically compute the bispectrum of primordial curvature perturbations and the corresponding non-Gaussianity parameter $f_{\mathrm{NL}}$ defined by \eqref{Fnl}.
One may wonder if our numerical results are accurate in spite of the complexity of the computation.
Fortunately, there is a consistency relation given by \cite{Maldacena:2002vr}
\begin{equation}\label{consistency}
   \frac{12}{5} \lim_{k_3\rightarrow 0} f_{\mathrm{NL}}(k_1,k_2,k_3)=1-n_{\mathrm s},\quad \text{for}~ k_1=k_2.
\end{equation}
It was shown that this relation always holds for single field inflation \cite{Creminelli:2004yq}, which can be an ideal tool to test our numerical results. 
We show the spectral index and the non-Gaussianity parameter $f_{\mathrm{NL}}$ in squeezed limit for the parameter set I with $p=1/2$ in Fig. \ref{fig:fnl}. 
From Fig. \ref{fig:fnl}, the consistency relation \eqref{consistency} holds, which enhances our confidence on the numerical results, more or less.
Besides, we also show the non-Gaussianity parameter $f_{\mathrm{NL}}$ in the equilateral limit. 
From Fig. \ref{fig:fnl}, the non-Gaussianity parameter is small at peak scales though it can be pretty large at the beginning of the USR phase.

\section{PBHs and SIGWs}\label{PBHGW}
Large perturbations will cause gravitational collapse to form PBHs after horizon reentry during radiation domination accompanied by the generation of SIGWs.
In this section, we compute the abundance of PBHs and the fractional energy density of SIGWs from the $E$ model. 
We also consider the effects of non-Gaussianity. 

\subsection{PBHs}

\begin{figure}[htp]
\centering
\subfigure{\includegraphics[width=0.45\linewidth]{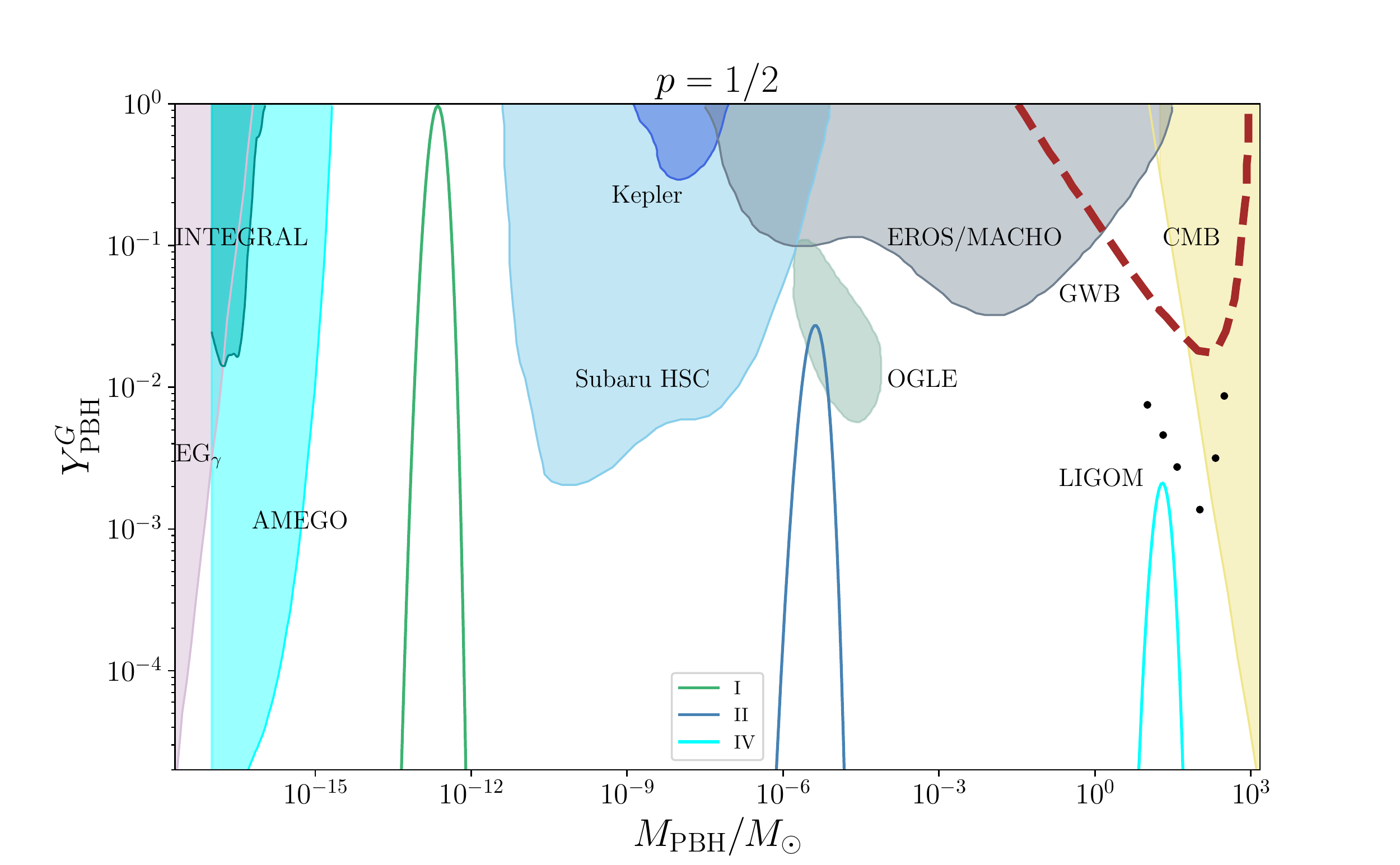}}
\subfigure{\includegraphics[width=0.45\linewidth]{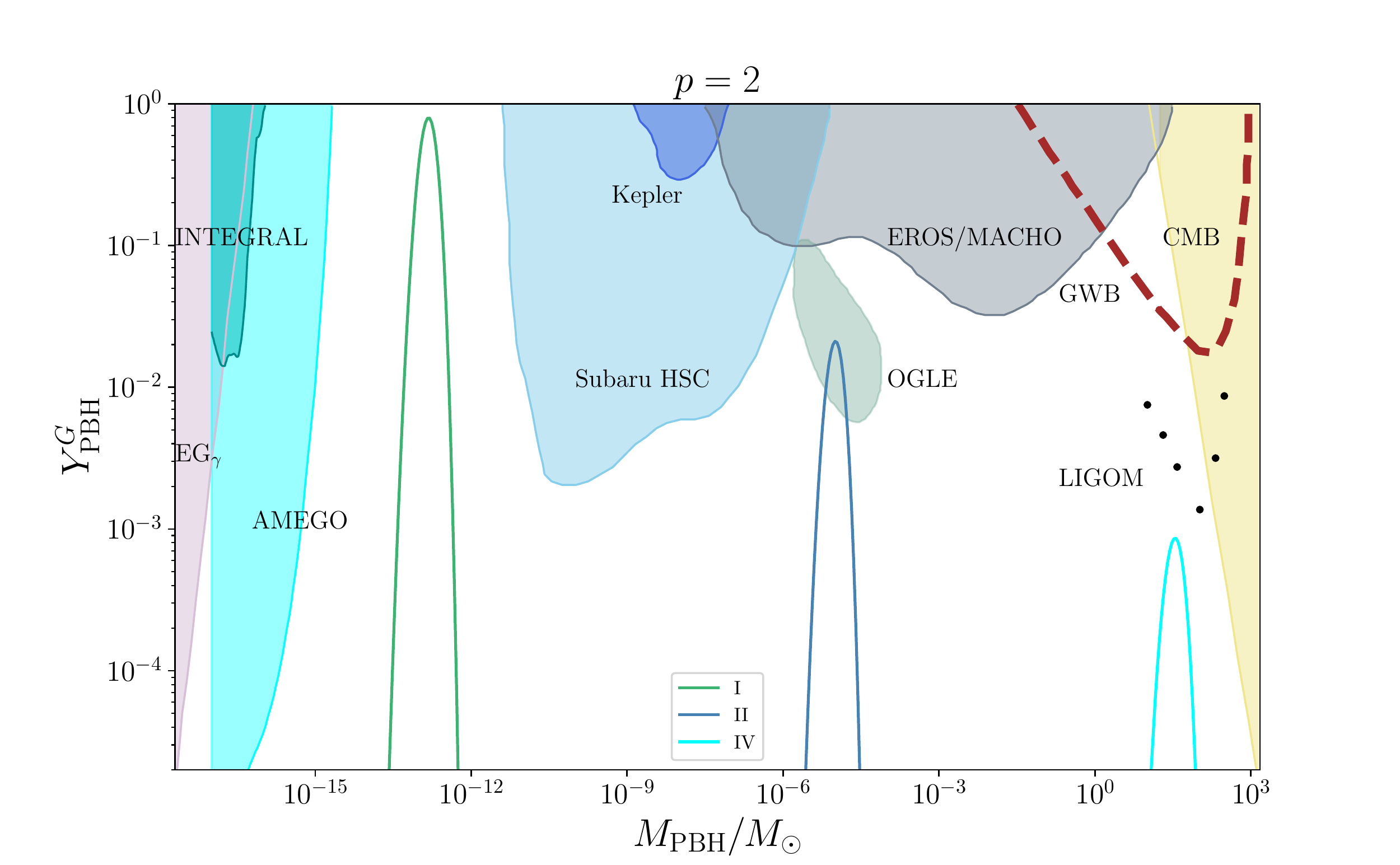}}
\caption{The PBH DM abundance $Y^G_{\mathrm{PBH}}$ produced by the $E$ model with Gaussian statistic perturbation.
The left panel shows the results for $p=1/2$, and the right panel shows the results for $p=2$.
The green-cyan shaded region presents the allowed PBH abundance from the ultrashort-timescale microlensing events in the OGLE data \cite{mroz2017no,Niikura:2019kqi}.
The other shaded regions, the dashed line, and the dotted line are the constraints on PBHs abundance from various observations \cite{Sato-Polito:2019hws,Carr:2009jm,Laha:2019ssq,Dasgupta:2019cae,Niikura:2017zjd,Griest:2013esa,Tisserand:2006zx,Ali-Haimoud:2017rtz,Raidal:2017mfl,Ali-Haimoud:2016mbv,Poulin:2017bwe,Wang:2019kaf,Laha:2020ivk,Laha:2020vhg}.}
\label{fig:PBHs}
\end{figure}

The mass of PBH at the formation time is $\gamma M_{\mathrm{hor}}$, where $M_{\mathrm{hor}}$ is the horizon mass and $\gamma$ is a numerical factor that depends on the details of gravitational collapse.
We choose the factor
$\gamma= 0.2$ \cite{Carr:1975qj}. 
The current fractional energy density of PBHs with mass $M$ in DM is \cite{Carr:2016drx,Gong:2017qlj}
\begin{equation}
\label{fpbheq1}
\begin{split}
Y_{\text{PBH}}(M)=&\frac{\beta(M)}{3.94\times10^{-9}}\left(\frac{\gamma}{0.2}\right)^{1/2}
\left(\frac{g_*}{10.75}\right)^{-1/4}\times \left(\frac{0.12}{\Omega_{\text{DM}}h^2}\right)
\left(\frac{M}{M_\odot}\right)^{-1/2},
\end{split}
\end{equation}
where $M_{\odot}$ is the solar mass, $g_*$ is the effective degrees of freedom at the formation time, $\Omega_{\text{DM}}$ is the current	energy density parameter of DM, and $\beta$ is the fractional energy density of PBHs at the formation.
In Press-Schechter theory \cite{Press:1973iz}, the fraction $\beta$ could be regarded as the probability that the density contrast exceeds the threshold
\begin{equation}\label{beta:pro}
  \beta=2\int_{\delta_c}^{\infty}P(\delta) d\delta,
\end{equation}
where $P(\delta)$ is the probability distribution function (PDF) of density contrast $\delta$, and $\delta_c$ is the threshold for the formation of PBHs.
For Gaussian statistic
\begin{equation}
\label{eq:beta}
\beta^G(M)\approx\sqrt{\frac{2}{\pi}}\frac{\sigma(M)}{\delta_c}
\exp\left(-\frac{\delta_c^2}{2\sigma^2(M)}\right),
\end{equation}
where $\sigma^2$ is the variance of $\delta$ smoothed on the horizon scales $R=1/(aH)$,
\begin{equation}
\label{sigmaeq1}
\sigma^2(M)=\left(\frac{4}{9}\right)^2\int \frac{dk}{k} \widetilde{\text{W}}^2(kR)(kR)^4\mathcal{P}_{\mathcal{R}}(k),
\end{equation}
We choose a Gaussian window function $\widetilde{\mathrm{W}}(x)=\mathrm{e}^{-x^2/2}$.
The effective degree of freedom is $g_*=107.5$ for $T>300$ GeV
and $g_*=10.75$ for $0.5\text{~MeV}<T<300\text{~GeV}$.
We take the observational value $\Omega_{\text{DM}}h^2=0.12$ \cite{Aghanim:2018eyx}
and threshold $\delta_c=0.44$ \cite{Harada:2013epa,Tada:2019amh,Escriva:2019phb}
in the calculation of PBH abundance.
The relation between the PBH mass $M$ and the scale $k$ is \cite{Gong:2017qlj}
\begin{equation}
\label{mkeq1}
M(k)=3.68\left(\frac{\gamma}{0.2}\right)\left(\frac{g_*}{10.75}\right)^{-1/6}
\left(\frac{k}{10^6\ \text{Mpc}^{-1}}\right)^{-2} M_{\odot}.
\end{equation}
With Eqs. \eqref{fpbheq1}, \eqref{eq:beta}, \eqref{mkeq1} and the power spectrum obtained in Sec. \ref{GB}, we compute the PBH DM abundance with Gaussian approximation, $Y^G_{\mathrm{PBH}}$, and the results are shown in Table \ref{results:tab} and Fig. \ref{fig:PBHs}.
For parameter set I, the $E$ model with both $p=1/2$ and $p=2$ produces PBHs with mass around $M_{\mathrm{PBH}}\sim 10^{-13}-10^{-12}M_{\odot}$.
The PBH abundance at the peak is $Y^G_{\mathrm{PBH}}\simeq 0.969$ for $p=1/2$ and  $Y^G_{\mathrm{PBH}}\simeq 0.79$ for $p=2$, respectively.
In this mass range, PBHs can be all DM.
For parameter set II, our model produces PBHs with the mass range $M_{\mathrm{PBH}}\sim 10^{-6}-10^{-5}M_{\odot}$ for both $p=1/2$ and $p=2$, the abundance of PBHs are $Y^G_{\mathrm{PBH}}\simeq 0.027$ and $Y^G_{\mathrm{PBH}}\simeq 0.021$, respectively. 
PBHs in this range can explain the ultrashort-timescale microlensing events in the OGLE data \cite{mroz2017no,Niikura:2019kqi}.
For parameter set IV, the models of both $p=1/2$ and $p=2$ produce PBHs with mass around $M_{\mathrm{PBH}}\sim 30M_{\odot}$.
These PBHs can explain the LIGO events.

\begin{figure}[htp]
\centering
\includegraphics[width=0.7\linewidth]{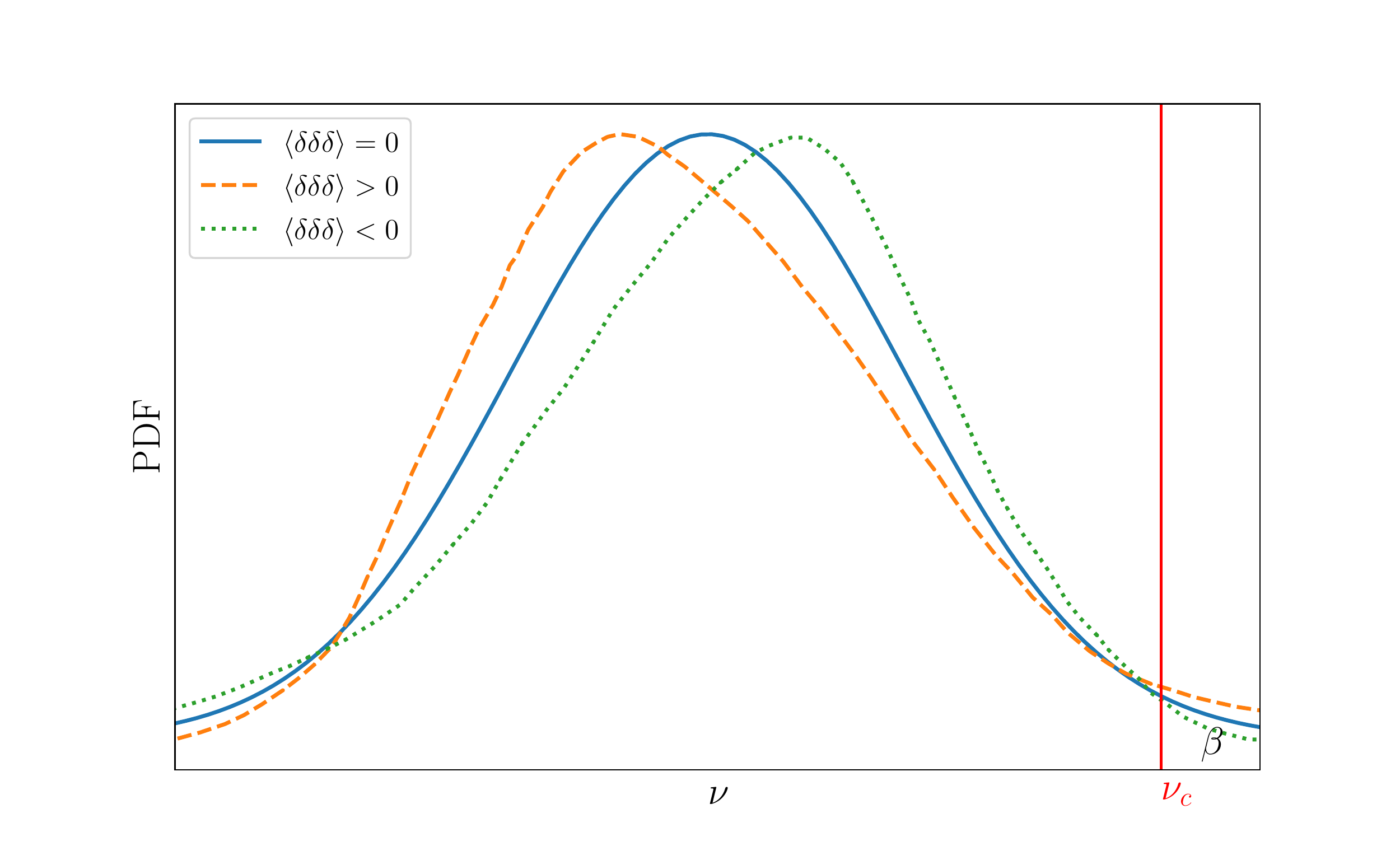}
\caption{The PDF of $\delta$ for $\langle \delta \delta \delta \rangle =0$ (solid line), $\langle \delta \delta \delta \rangle >0$ (dashed line), and $\langle \delta \delta \delta \rangle <0$ (dotted line). The red vertical line represents the threshold for PBH formation. The area below the PDF with $\nu\geq \nu_c$ represents the PBH abundance at the formation time.}
\label{fig:PDF}
\end{figure}

We discuss the role that the non-Gaussianity plays in the formation of PBHs qualitatively, and then we present the numerical result.
From the viewpoint of statistics, a non-vanishing three-point function that relates to the skewness of a PDF implies a non-Gaussian tail, as the schematic shown in Fig. \ref{fig:PDF} for illustration.
A positive three-point function, $\langle \delta \delta \delta \rangle>0$, corresponds to a right-tailed PDF, so the probability of obtaining perturbations with $\nu=\delta/\sigma\gg 1$ becomes larger relative to a Gaussian PDF, and vice versa.
The formation of PBHs is a rare event, mainly attributed to the large perturbations with $\nu\geq\nu_c=\delta_c/\sigma\gg 1$.
The abundance of PBHs $\beta$ at the formation corresponds to the area below the PDF with $\nu\geq \nu_c$ according to Eq. \eqref{fig:PDF}.
This is sensitive to the non-Gaussian tail.
In addition, from Fig. \ref{fig:PDF}, we see that the Gaussian-predicted PBH abundance $\beta^G$ is enhanced/suppressed if the three-point function is positive/negative.

Then we discuss the effect of non-Gaussianity on the PBH abundance, quantitatively. 
The mass fraction $\beta$ with non-Gaussian corrected \cite{Franciolini:2018vbk,Riccardi:2021rlf}
\begin{equation}
    \beta=\text{e}^{\Delta_3}\beta^G,
\end{equation}
where the 3rd cumulant $\Delta_3$
\begin{equation}\label{delta3}
\Delta_3=\frac{1}{3!}\left(\frac{\delta_c}{\sigma}\right)^2 \mathcal{S}_3 \delta_c,
\end{equation}
with
\begin{equation}
\mathcal{S}_3=\frac{\left\langle\delta(\bm{x}) \delta(\bm{x}) \delta(\bm{x})\right\rangle}{\sigma^4}.
\end{equation}
Since the mass of PBHs is almost monochromatic, we only consider the correction from peak-scale perturbation \cite{Zhang:2021vak}
\begin{equation}\label{C3}
\Delta_3\approx  23\frac{\delta^3_c}{\mathcal{P}_{\mathcal{R}}(k_{\text{peak}})}f_{\mathrm{NL}}(k_{\text{peak}},k_{\text{peak}},k_{\text{peak}}).
\end{equation}

We compute $f_{\mathrm{NL}}(k_{\text{peak}},k_{\text{peak}},k_{\text{peak}})$ numerically and show the results in Table. \ref{results:tab}. The non-Gaussianity parameter $f_{\mathrm{NL}}(k_{\text{peak}},k_{\text{peak}},k_{\text{peak}})$ is about $\mathcal{O}(0.1)$, and $\Delta_3 \sim \mathcal{O}(10)$, which means the PBH abundance is highly underestimated using just Gaussian approximation.
There are more PBHs due to the right-tailed PDF, which is consistent with our above qualitative discussion.

\subsection{SIGWs}
The energy density parameter of SIGWs per logarithmic interval of ${k}$ can be expressed as following \cite{Kohri:2018awv,Espinosa:2018eve}
\begin{equation}\label{EGW0}
\Omega_{\mathrm{GW}}\left(k\right)=\frac{1}{24}\left(\frac{k}{a(\tau)H(\tau)}\right)^2\overline{\mathcal{P}_h\left(k,\tau\right)},
\end{equation}
where the power spectrum of SIGWs averaged for several wavelengths $\overline{\mathcal{P}_h}$
\begin{equation}\label{Eph2}
\overline{\mathcal{P}_{h}\left(k,\tau\right)}=4\int^{\infty}_{0}dv\int^{1+v}_{\left|1-v\right|}du \left(\frac{4v^2-\left(1+v^2-u^2\right)^2}{4vu}\right)^2
\overline{I^{2}_{\mathrm{RD}}\left(u,v,x\right)}\mathcal{P}_\mathcal{R}\left(uk\right)\mathcal{P}_\mathcal{R}\left(vk\right),
\end{equation}
with $x=k\tau$ and ${I^{2}_{\mathrm{RD}}\left(u,v,x\rightarrow\infty \right)}$ is
\begin{equation}\label{IRD}
\begin{split}
\overline{I_{\mathrm{RD}}^2(v,u,x\rightarrow \infty)} =& \frac{1}{2x^2} \left( \frac{3(u^2+v^2-3)}{4 u^3 v^3 } \right)^2 \left\{ \left( -4uv+(u^2+v^2-3) \log \left| \frac{3-(u+v)^2}{3-(u-v)^2} \right| \right)^2  \right. \\
&  \left.\vphantom{\log \left| \frac{3-(u+v)^2}{3-(u-v)^2} \right|^2} \qquad \qquad \qquad \qquad \qquad   + \pi^2 (u^2+v^2-3)^2 \Theta \left( v+u-\sqrt{3}\right)\right\}. 
\end{split}
\end{equation}
The fractional energy density of SIGWs today $\Omega_{\mathrm{GW},0}$ 
\cite{Espinosa:2018eve}
\begin{equation}\label{EGW}
\Omega_{\mathrm{GW},0}\left(k\right)=\Omega_{\mathrm{GW}}\left(k\right)\frac{\Omega_{r,0}}{\Omega_{r}\left(\tau\right)},
\end{equation}
where $\Omega_{r,0}$ is the current fractional energy density of radiation. We choose $\Omega_{r}=1$ during the radiation domination.

\begin{figure}[htp]
\centering
\subfigure{\includegraphics[width=0.45\linewidth]{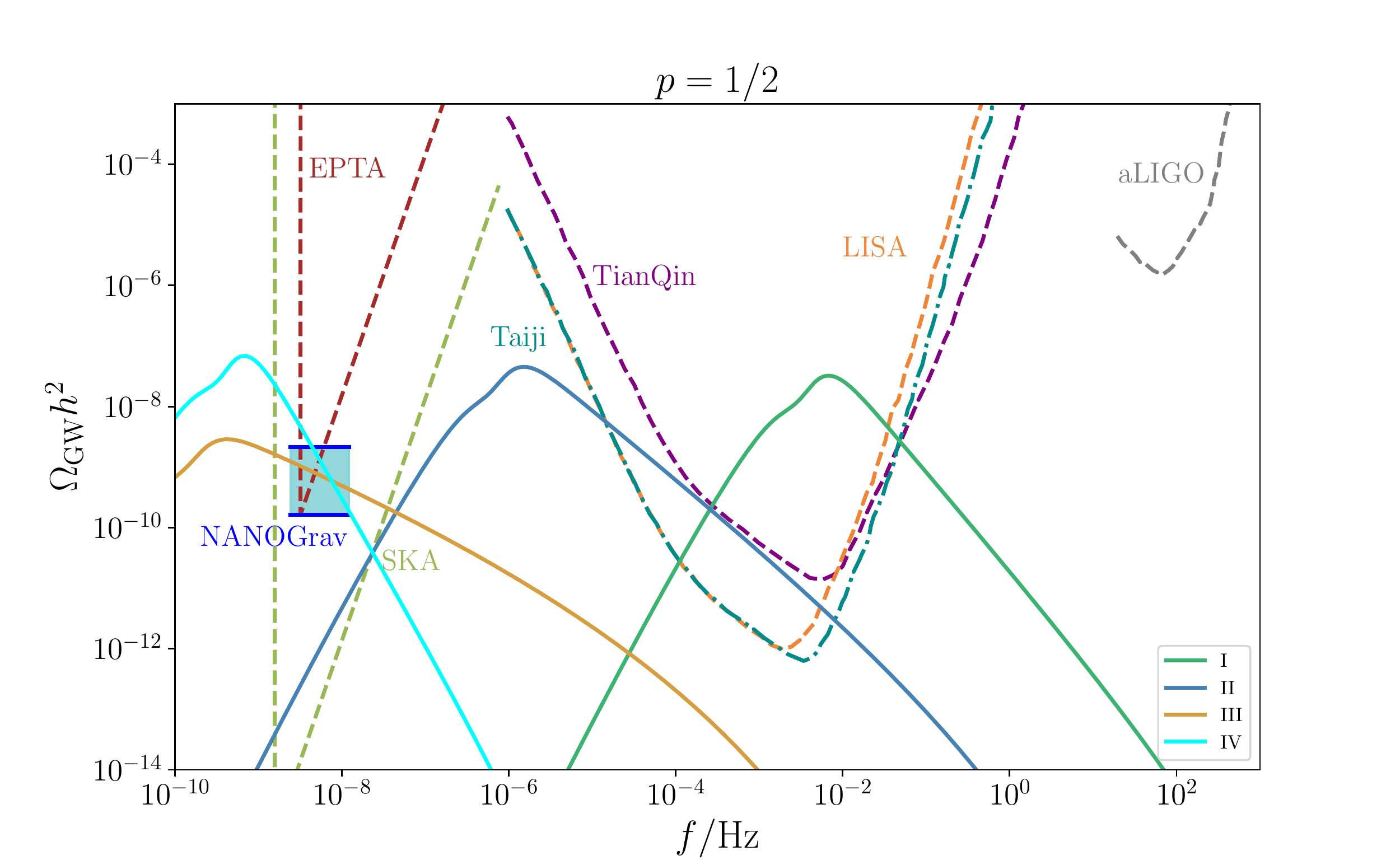}}
\subfigure{\includegraphics[width=0.45\linewidth]{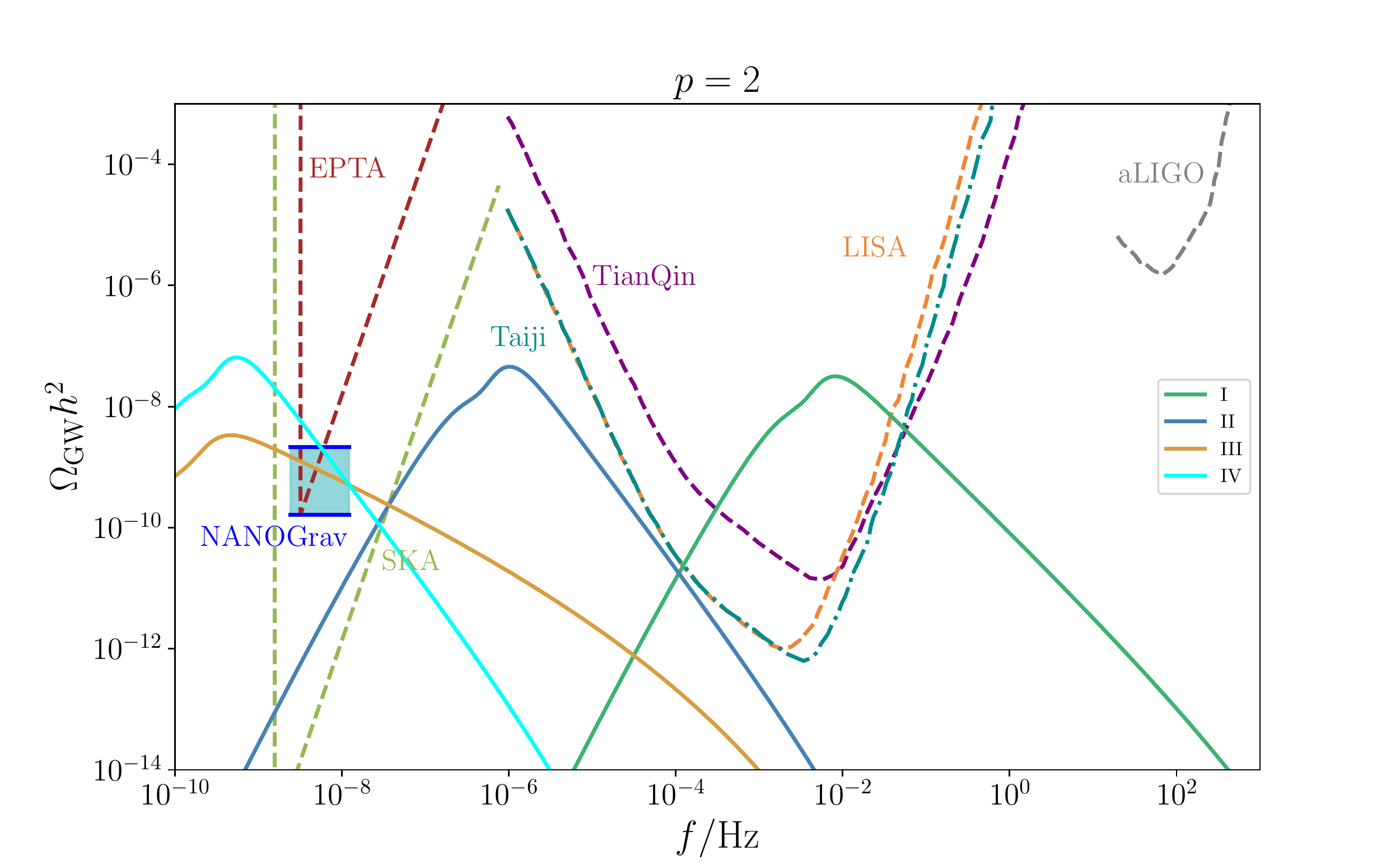}}
\caption{The SIGWs generated by the $E$ model with $p=1/2$ (left panel) and $p=2$ (right panel). The solid curves represent the energy density of SIGWs. The fuchsia dashed curve shows the EPTA limit \cite{Ferdman:2010xq,Hobbs:2009yy,McLaughlin:2013ira,Hobbs:2013aka} ,
the SKA limit \cite{Moore:2014lga} is shown in the lime dashed curve, the purple dashed curve in the middle denotes the TianQin limit \cite{Luo:2015ght}, the dark cyan dotted-dashed curve shows the Taiji limit \cite{Hu:2017mde}, the orange dashed curve denotes the LISA limit \cite{Audley:2017drz}, and the aLIGO limit \cite{Harry:2010zz,TheLIGOScientific:2014jea} shown in the gray dashed curve.}
\label{fig:SIGWs}
\end{figure}

With the power spectrum obtained in Sec. \ref{GB}, we compute the fractional energy density of SIGWs from the $E$ model, the results are shown in Table. \ref{results:tab} and Fig. \ref{fig:SIGWs}.
The SIGWs with peak frequency being milli-Hertz can be detected by TianQin, Taiji, and LISA. 
The SIGWs with peak frequency $f_c\sim 10^{-6}$Hz can be tested by SKA. 
For parameter set IV, the SIGWs can be observed by both SKA and EPTA.
With parameter set III, the $E$ model produces a broad power spectrum, and the corresponding spectrum of SIGWs is flatter than the others.
The energy density of SIGWs lies within the $2\sigma$ region of the NANOGrav signal \cite{DeLuca:2020agl,Vaskonen:2020lbd,Kohri:2020qqd,Domenech:2020ers,Vagnozzi:2020gtf}. 

To estimate the contribution from non-Gaussianity, we use nonlinear coupling constant $f_{\text{NL}}$ defined by \cite{Verde:1999ij,Komatsu:2001rj}
\begin{equation}\label{local}
\mathcal{R}(\bm{x})=\mathcal{R}^G(\bm{x})+\frac{3}{5}f_\text{NL}(\mathcal{R}^G(\bm{x})^2-\langle \mathcal{R}^G(\bm{x})^2 \rangle),
\end{equation}
where $\mathcal{R}^G$ is the linear Gaussian part of the curvature perturbations.
Then the power spectrum of curvature perturbations taking into account the non-Gaussianity can be expressed as
\begin{equation}\label{sumps}
\mathcal{P}_{\mathcal{R}}(k)=\mathcal{P}^G_{\mathcal{R}}(k)+\mathcal{P}^{NG}_{\mathcal{R}}(k),
\end{equation}
where
\begin{equation}
\mathcal{P}^{NG}_{\mathcal{R}}(k)
=\left(\frac{3}{5}\right)^2\frac{k^3}{2\pi}f^2_{\mathrm{NL}}
\int d^3\bm{p}\frac{\mathcal{P}^G_{\mathcal{R}}(p)}{p^3}
\frac{\mathcal{P}^G_{\mathcal{R}}(|\bm{k}-\bm{p}|)}{|\bm{k}-\bm{p}|^3}.
\end{equation} 

The non-Gaussian contribution to the energy density of SIGWs is insignificant unless $\left(\frac{3}{5}\right)^2f^2_{\text{NL}} \mathcal{P}^G_\mathcal{R} \gtrsim 1$ \cite{Cai:2018dig}.
From our results, $f_{\text{NL}}\sim \mathcal{O}(0.1)$, and $\mathcal{P}^G_{\mathcal{R}}\sim \mathcal{O}(0.01)$ at peak scales, so the effects of non-Gaussianity on SIGWs is neglegible.

\section{Conclusion}\label{conclusion}
The higher-order curvature correction of gravity may play an important role in the early Universe. 
In this paper, we use an $E$ model potential to drive inflation with the Gauss-Bonnet term. 
In this model, the power spectrum can not only satisfy the constraints from CMB on large scales, but also reach $\mathcal{O}(0.01)$ on small scales due to Gauss-Bonnet coupling.
The large perturbations cause gravitational collapse to form PBHs, accompanied by the generation of SIGWs after horizon reentry during the radiation domination. 
By tuning the model parameters, the power spectrum can be enhanced at various scales and thus we can obtain PBHs with different mass ranges.
The PBHs with mass range $\mathcal{O}(10^{-13})-\mathcal{O}(10^{-12})M_{\odot}$ makes up almost all DM, and the corresponding SIGWs is observable by space-based GWs observatories, like TianQin, Taiji and LISA.
The abundance of PBHs with mass in the range
$\mathcal{O}(10^{-6})-\mathcal{O}(10^{-5})M_{\odot}$ can be $Y^{G}_{\mathrm{PBH}}\sim \mathcal{O}(0.02)$.
These PBHs could explain the ultrashort-timescale microlensing events in the OGLE data, and the SIGWs can be tested by SKA.
With the parameter set IV, the mass of PBHs is around $\sim 30M_{\odot}$.
These PBHs could explain the LIGO-Virgo events.
Meanwhile, the Nanohertz SIGWs are observable by SKA and EPTA.
With parameter set III, we get a broad peak in the power spectrum, and the corresponding SIGWs can explain the signal hinted by NANOGrav.

We also consider the effect of non-Gaussianity on the abundance of PBHs and the fractional energy density of SIGWs.
We find that more PBHs can be produced.
The abundance of PBHs receives a significant enhancement due to the right-tailed PDF. 
However, the fractional energy density of SIGWs is insensitive to non-Gaussianity and remains almost unaffected.

\begin{acknowledgments}
The author would like to thank Jiong Lin and Yizhou Lu for useful discussion. This work was partly supported by the National Natural Science Foundation of China (NSFC) under the grant No.
11975020.
\end{acknowledgments}

\appendix 
\section{The Cubic Action And The Bispectrum}
\label{app.NG}
The cubic action used to compute the bispectrum is \cite{DeFelice:2011zh}
\begin{equation}\label{S3}
\begin{split}
S_{3}=& \int dt d^{3} x\left\{a^{3} \mathcal{C}_{1}\mathcal{R} \dot{\mathcal{R}}^{2}+a \mathcal{C}_{2} \mathcal{R}(\partial \mathcal{R})^{2}+a^{3} \mathcal{C}_{3}\dot{\mathcal{R}}^{3}+a^{3} \mathcal{C}_{4} \dot{\mathcal{R}}\left(\partial_{i} \mathcal{R}\right)\left(\partial_{i} \chi\right)\right.\\
&+a^{3}\mathcal{C}_{5}\partial^{2} \mathcal{R}(\partial \chi)^{2}+a \mathcal{C}_{6} \dot{\mathcal{R}}^{2} \partial^{2} \mathcal{R}+\left(\mathcal{C}_{7} / a\right)\left[\partial^{2} \mathcal{R}(\partial \mathcal{R})^{2}-\mathcal{R} \partial_{i} \partial_{j}\left(\partial_{i} \mathcal{R}\right)\left(\partial_{j} \mathcal{R}\right)\right] \\
&\left.+a\mathcal{C}_{8} \left[\partial^{2} \mathcal{R} \partial_{i} \mathcal{R} \partial_{i} \chi-\mathcal{R} \partial_{i} \partial_{j}\left(\partial_{i} \mathcal{R}\right)\left(\partial_{j} \chi\right)\right]+\left.\mathcal{F}_{1} \frac{\delta \mathcal{L}_{2}}{\delta \mathcal{R}}\right|_{1}\right\}
\end{split}
\end{equation}
where $\mathcal{L}_2$ is the quadratic Lagrangian and $\partial^2\mathcal{\chi}=Q\dot{\mathcal{R}}$, and 
\begin{equation}
\begin{split}
\mathcal{F}_1=& \frac{A_5}{w_{1}^{2}}\left\{\left(\partial_i \mathcal{R}\right)\left(\partial_i \chi\right)-\partial^{-2} \partial_{i} \partial_{j}\left[\left(\partial_{i} \mathcal{R}\right)\left(\partial_{j} \chi\right)\right]\right\} \\
&+q_{1} \mathcal{R} \dot{\mathcal{R}}+\frac{A_{7}-2 A_{5} L_{1}}{4 w_{1} a^{2}}\left\{(\partial \mathcal{R})^{2}-\partial^{-2} \partial_{i} \partial_{j}\left[\left(\partial_{i} \mathcal{R}\right)\left(\partial_{j} \mathcal{R}\right)\right]\right\}.
\end{split}
\end{equation}
The expressions of  $\mathcal{C}_1-\mathcal{C}_8$ are as follows
\begin{equation}
\begin{split}
    &w_1=1-\frac{1}{2}\xi_1,\ \ w_2=2H(1-\frac{3}{4}\xi_1),
    \\& w_3=-3H^2(3-\epsilon_1-\frac{11}{4}\xi_1+\frac{1}{4}\epsilon_1\xi_1-\frac{1}{4}\xi_1\xi_2),
    \\&w_4=1-\frac{1}{2}(\epsilon_1\xi_1+\xi_1\xi_2),\ \ L_1=\frac{2w_1}{w_2};
\end{split}
\end{equation}
\begin{equation}
\begin{split}
    &a_1=H^2(3-\epsilon_1-\frac{19}{4}\xi_1+\frac{1}{4}\epsilon_1\xi_1-\frac{1}{4}\xi_1\xi_2),\ \ a_2=w_3,
    \\& a_3=-3H(2-3\xi_1),\ \ a_4=-\xi_1,\ \ a_5=H(2-3\xi_1),
    \\& a_6=-w_2,\ \ a_7=9w_2,\ \ a_8=\frac{\xi_1}{H}, \ \ a_9=-\frac{1-3\xi_1/2}{2},
    \\& a_{10}=\frac{\xi_1}{2H},\ \ a_{11}=-w_2,\ \ a_{12}=4a_9,\ \ a_{13}=-2w_1,
    \\& a_{14}=-w_1,\ \  a_{15}=-6a_9;
\end{split}
\end{equation}
\begin{equation}
    b_1=\frac{1}{2}a_8,\ \ b_2=w_4,\ \ b_3=-9w_1;
\end{equation}
\begin{equation}
    c_1=c_3=-a_{13},\ \ c_2=-a_{10};
\end{equation}
\begin{equation}
    d_1=-\frac{\xi_1}{4H},\ \ d_2=\frac{3}{2}w_1,\ \ d_3=-2w_1;
\end{equation}

\begin{equation}
\begin{split}
&A_{1}=b_{1}+L_{1} a_{15}+L_{1}^{2} a_{3}+L_{1}^{3} a_{1}, \ \ A_{2}=L_{1}\left(L_{1} a_{4}+a_{8}\right),\\
&A_{3}=c_{2}+L_{1} a_{12}+L_{1}^{2} a_{5},\ \  A_{4}=b_{3}+L_{1} a_{7}+L_{1}^{2} a_{2}, \\
&A_{5}=L_{1} a_{9}+d_{1}, \ \ A_{6}=d_{2},\ \ A_{7}=L_{1} a_{10}, \\
&A_{8}=b_{2}+a_{13} \dot{L}_{1} / 2+L_{1}\left(\dot{a}_{13}+H a_{13}\right)/2,\ \ A_{9}=d_{3};
\end{split}
\end{equation}

\begin{equation}
\begin{split}
& q_1=-\frac{L_1}{c^2_s},\\
& q_{2}=-\frac{4 A_{6} L_{1}}{w_{1}}-a^{2} \frac{d}{d t}\left(\frac{A_{7}-2 A_{5} L_{1}}{a^{2} w_{1}}\right)-\frac{2 A_{9} L_{1}}{w_{1}},\\
& q_{3}=A_{6} L_{1}^{2}-\frac{a}{3} \frac{d}{d t}\left(\frac{A_{5} L_{1}^{2}-A_{7} L_{1}}{a}\right)+\frac{2}{3} A_{9} L_{1}^{2}.
\end{split}
\end{equation}

\begin{equation}
\begin{split}
&\mathcal{C}_{1} =A_{4}+q_{1}(\dot{Q}+3 H Q)-Q \dot{q}_{1}, \ \ \mathcal{C}_{2} =A_{8}+\frac{1}{a}\frac{d}{d t}\left(a L_{1} Q\right), \\
&\mathcal{C}_{3} =A_{1}+A_{3} \frac{Q}{w_{1}}-q_{1} Q, \\
&\mathcal{C}_{4} =\frac{Q}{w_{1}}\left[\frac{1}{w_{1}}\left(A_{6}+A_{9}\right)-w_{1} \frac{d}
{d t}\left(\frac{A_{5}}{w_{1}^{2}}\right)+\frac{3 H A_{5}}{w_{1}}\right],\\
&\mathcal{C}_{5}= \frac{1}{2}\left[\frac{A_{6}}{w_{1}^{2}}-\frac{d}{d t}\left(\frac{A_{5}}{w_{1}^{2}}\right)+\frac{3 H A_{5}}{w_{1}^{2}}\right], \\
&\mathcal{C}_{6}=A_{2}-A_{3} L_{1}, \ \ \mathcal{C}_{7}=q_{3}-\frac{Q c_{s}^{2}}{2 w_{1}}\left(A_{7}-2 A_{5} L_{1}\right), \ \ \mathcal{C}_{8}=\frac{q_{2}}{2}-\frac{2 c_{s}^{2} A_{5} Q}{w_{1}^{2}}.
\end{split}
\end{equation}

With the above expressions of the coefficients, the explicit form of bispectrum $B_\mathcal{R}$
\begin{equation}
 B_\mathcal{R}(k_1,k_2,k_3)=\mathcal{R}_{k_1}(t_e)\mathcal{R}_{k_2}(t_e)\mathcal{R}_{k_3}(t_e)\sum_{i=1}^{10}\mathcal{B}^i_\mathcal{R}(k_1,k_2,k_3),
\end{equation}
where
\begin{equation}
\mathcal{B}^1_\mathcal{R}(k_1,k_2,k_3)= -4\Im\int^{t_e}_{t_i} dta^3\mathcal{C}_1\left(\mathcal{R}^*_{k_1}(t)\dot{\mathcal{R}^*}_{k_2}(t)\dot{\mathcal{R}^*}_{k_3}(t)+\text{perm}\right),
\end{equation}

\begin{equation}
\mathcal{B}^2_\mathcal{R}(k_1,k_2,k_3)= 4\Im\int^{t_e}_{t_i} dt a\mathcal{C}_2\left[\left(\bm{k}_1\cdot\bm{k}_2+\bm{k}_1\cdot\bm{k}_3+\bm{k}_2\cdot\bm{k}_3\right)\mathcal{R}^*_{k_1}(t)\mathcal{R}^*_{k_2}(t)\mathcal{R}^*_{k_3}(t)\right],
\end{equation}

\begin{equation}
\mathcal{B}^3_\mathcal{R}(k_1,k_2,k_3)= -12\Im\int^{t_e}_{t_i} dt a^3\mathcal{C}_3\dot{\mathcal{R}^*}_{k_1}(t)\dot{\mathcal{R}^*}_{k_2}(t)\dot{\mathcal{R}^*}_{k_3}(t),
\end{equation}

\begin{equation}
\mathcal{B}^4_\mathcal{R}(k_1,k_2,k_3)= -2\Im\int^{t_e}_{t_i} dt a^3\mathcal{C}_4Q\left[\left(\frac{\bm{k}_1\cdot\bm{k}_2}{k^2_2}+\frac{\bm{k}_1\cdot\bm{k}_3}{k^2_3}\right)\mathcal{R}^*_{k_1}(t)\dot{\mathcal{R}^*}_{k_2}(t)\dot{\mathcal{R}^*}_{k_3}(t)+\text{perm}\right],
\end{equation}

\begin{equation}
\mathcal{B}^5_\mathcal{R}(k_1,k_2,k_3)= -4\Im\int^{t_e}_{t_i} dt a^3\mathcal{C}_5Q^2\left[\frac{k^2_1\bm{k}_2\cdot\bm{k}_3}{k^2_2k^2_3}\mathcal{R}^*_{k_1}(t)\dot{\mathcal{R}^*}_{k_2}(t)\dot{\mathcal{R}^*}_{k_3}(t)+\text{perm}\right],
\end{equation}

\begin{equation}
\mathcal{B}^6_\mathcal{R}(k_1,k_2,k_3)= 4\Im\int^{t_e}_{t_i} dt a\mathcal{C}_6\left[k^2_1\mathcal{R}^*_{k_1}(t)\dot{\mathcal{R}^*}_{k_2}(t)\dot{\mathcal{R}^*}_{k_3}(t)+\text{perm}\right],
\end{equation}

\begin{equation}
\mathcal{B}^7_\mathcal{R}(k_1,k_2,k_3)= -4\Im\int^{t_e}_{t_i} dt \mathcal{C}_7/a\left[\left(k^2_1\bm{k}_2\cdot\bm{k}_3+k^2_2\bm{k}_1\cdot\bm{k}_3+k^2_3\bm{k}_1\cdot\bm{k}_2\right)\mathcal{R}^*_{k_1}(t)\mathcal{R}^*_{k_2}(t)\mathcal{R}^*_{k_3}(t)\right],
\end{equation}

\begin{equation}
\begin{split}
\mathcal{B}^8_\mathcal{R}(k_1,k_2,k_3)= 2\Im\int^{t_e}_{t_i} dt \mathcal{C}_7/a &\left[\left(k^2_2\bm{k}_2\cdot\bm{k}_3+k^2_3\bm{k}_3\cdot\bm{k}_2+k^2_1\bm{k}_1\cdot\bm{k}_2+k^2_1\bm{k}_1\cdot\bm{k}_3 
\right.\right. \\ & \left.\left.
+k^2_2\bm{k}_2\cdot\bm{k}_1+k^2_3\bm{k}_3\cdot\bm{k}_1\right)\mathcal{R}^*_{k_1}(t)\mathcal{R}^*_{k_2}(t)\mathcal{R}^*_{k_3}(t)\right],
\end{split}
\end{equation}

\begin{equation}
\mathcal{B}^9_\mathcal{R}(k_1,k_2,k_3)= 2\Im\int^{t_e}_{t_i} dt a\mathcal{C}_8Q\left[\left(\frac{k^2_1\bm{k}_2\cdot\bm{k}_3}{k^2_3}+\frac{k^2_2\bm{k}_1\cdot\bm{k}_3}{k^2_3}\right)\mathcal{R}^*_{k_1}(t)\mathcal{R}^*_{k_2}(t)\dot{\mathcal{R}^*}_{k_3}(t)+\text{perm}\right],
\end{equation}

\begin{equation}
\mathcal{B}^{10}_\mathcal{R}(k_1,k_2,k_3)= -2\Im\int^{t_e}_{t_i} dt a\mathcal{C}_8Q\left[\left(\frac{k^2_2\bm{k}_2\cdot\bm{k}_3}{k^2_3}+\frac{k^2_1\bm{k}_1\cdot\bm{k}_3}{k^2_3}\right)\mathcal{R}^*_{k_1}(t)\mathcal{R}^*_{k_2}(t)\dot{\mathcal{R}^*}_{k_3}(t)+\text{perm}\right].
\end{equation}

The contribution of the last term in Eq. \eqref{S3} is negligible relative to those coming from other terms, as in Refs. \cite{DeFelice:2011uc,Arroja:2011yj,Rigopoulos:2011eq}, we neglect this term when computing the bispectrum.


%

\end{document}